\documentclass[onecolumn,floatfix,superscriptaddress,showpacs,showkeys,nofootinbib,preprint]{revtex4}
\usepackage{epsfig}
\usepackage{amssymb,latexsym,amsmath}
\newcommand{\eq}[1]{\begin{align} #1 \end{align}}

\begin{document}
\title{ Equation of State for the Quark Gluon Plasma\\
with the Negative Bag Constant
}

 \author{Viktor V. Begun}
 \affiliation{Bogolyubov Institute for Theoretical Physics,
 Kiev, Ukraine}
 \affiliation{Frankfurt Institute for Advanced Studies, Frankfurt,
 Germany}

 \author{Mark I. Gorenstein}
 \affiliation{Bogolyubov Institute for Theoretical Physics,
 Kiev, Ukraine}
 \affiliation{Frankfurt Institute for Advanced Studies, Frankfurt, Germany}

 \author{Oleg A. Mogilevsky}
 \affiliation{Bogolyubov Institute for Theoretical Physics,
 Kiev, Ukraine}

\begin{abstract}
We propose a new analytical  form of the quark-gluon plasma
equation of state (EoS). This new EoS satisfies all qualitative
features observed in the lattice QCD calculations and gives a good
quantitative description of the lattice results in the SU(3)
gluodynamics. The energy density for the suggested EoS looks
similar to that in the bag model, but requires a negative value of
the bag constant.
 \end{abstract}



\maketitle

The transition from a confined hadron-resonance phase to a
deconfined phase, the quark gluon plasma (QGP), is expected at
high temperatures and/or baryonic chemical potentials.  For
several decades,   the bag model (BM) equation of state (EoS) has
been used to describe the QGP (see, e.g., Ref.~\cite{bag}). In the
simplest form, i.e. for non-interacting massless constituents and
zero values of all conserved  charges, the BM EoS reads:
\eq{\label{bag}
\varepsilon(T)~=~\sigma_{SB}~T^4~+~B~,~~~~~~p(T)~=~\frac{\sigma_{SB}}{3}~T^4~-~B~,
}
where the energy density $\varepsilon$ and the pressure $p$  have
a simple dependence on temperature $T$ modified by adding a
positive constant $B$ (``vacuum pressure''). The Stefan-Boltzmann
(SB) constant $\sigma_{SB}$ in Eq.~(\ref{bag}) equals to:
\eq{\label{sigma}
\sigma_{SB}~=~
\frac{\pi^2}{30}~ \left(
d_B~+~\frac{7}{8}d_F\right) ~,
}
where $d_B$ and $d_F$ are the degeneracy factors for the bosons
(gluons) and fermions (quarks and antiquarks), respectively. The
zero value of the baryonic chemical potential in Eq.~(\ref{bag})
is a valid approximation for the QGP created  in nucleus-nucleus
collisions at the BNL RHIC and even better for future experiments
at the CERN LHC. Note that also most lattice QCD calculations for
the QGP EoS correspond to zero or very small values of the
baryonic chemical potential.  Equation (\ref{bag}) is assumed to
be valid at $T>T_c$, where the critical temperature $T_c$
corresponds to a  $1^{\text{st}}$  order phase transition in the
pure SU(3) gluodynamics or to a smooth crossover in the full QCD.

\begin{figure}[ht!]
\epsfig{file=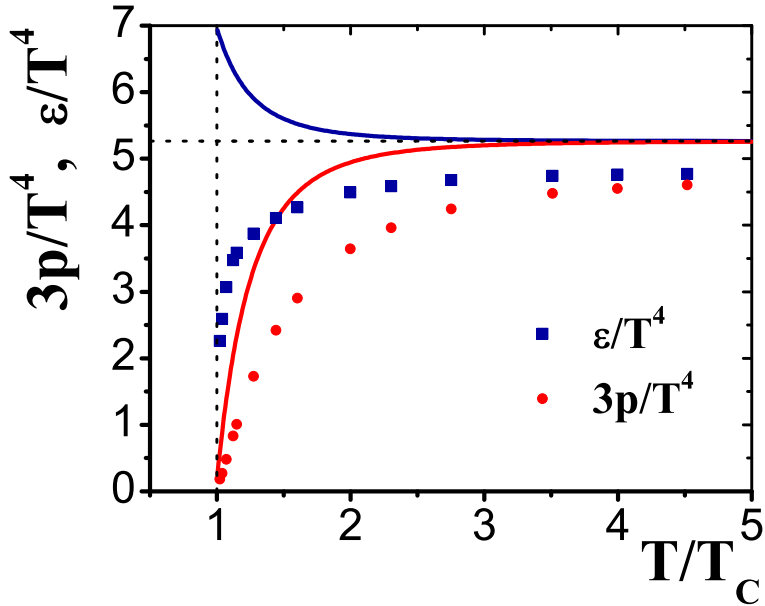,width=0.6\textwidth}
\caption{ The MC lattice results for SU(3) gluodynamics
\cite{lattice} for the energy density  (squares) and pressure
(circles) at $T> T_c$. The size of the symbols corresponds to the
error-bars reported in Ref.~\cite{lattice}. The dashed horizontal
line corresponds to the SB constant $\sigma_{SB}=8\pi^2/15$~. The
solid lines show the BM EoS (\ref{bag}) with $d=16$ and $B= 1.7
T_c^4$ for $\varepsilon/T^4$ (upper line) and $3p/T^4$ (lower
line). \label{fig:MC-data}}
\end{figure}

The main features of the QCD deconfined matter EoS can be
illustrated by the Monte Carlo (MC) lattice results~\cite{lattice}
for the SU(3) gluodynamics presented in Fig.~\ref{fig:MC-data}.
 They can be qualitatively summarized as follows:

1. The pressure $p(T)$ rapidly increases at $T\gtrsim T_c$, while
at high $T$ the system reaches the ideal massless gas behavior $p
\cong \varepsilon/3$.
%
%

2.  However, the constant $\sigma\cong \varepsilon/T^4\cong
3p/T^4$ observed at high $T$ is about $10\div 20\%$ smaller than
the value of $\sigma_{SB}$ in Eq.~(\ref{sigma}).

3. At high $T$, both $\varepsilon/T^4$ and $3p/T^4$ approach their
limiting value $\sigma $ from below.
%

Note that  these properties of the gluon plasma EoS are also valid
in the full QCD.

\vspace{0.2cm}
The phenomenological success of the BM EoS (\ref{bag}) is due to
the fact that it satisfies the first property: it gives $p\cong
\varepsilon/3$ at high $T$ and shows an abrupt drop of $p(T)$ near
$T_c$. However, the BM EoS is in a contradiction with the second
and third features listed above (see Fig.~\ref{fig:MC-data}). The
BM EoS (\ref{bag}) gives no suppression of the SB constant. Note
that we restrict our consideration to the present lattice results
available at finite temperature interval $T_c<T<4.5T_c$ and do not
discuss the possible asymptotic behavior at $T\rightarrow\infty$.
The BM energy density $\varepsilon(T)/T^4$ approaches its SB limit
from above. This contradicts the MC lattice results.  Despite
these evident problems, the BM EoS (\ref{bag}), due to its
simplicity, is still one of the most popular models for
phenomenological applications.
%

\vspace{0.2cm}
 In this letter we suggest a new analytical
parametrization for the QGP EoS. It satisfies all three properties
listed above, gives a good quantitative  description of the MC
lattice results for the SU(3) gluodynamics, and is almost as
simple as Eq.~(\ref{bag}).

\begin{figure}[ht!]
\epsfig{file=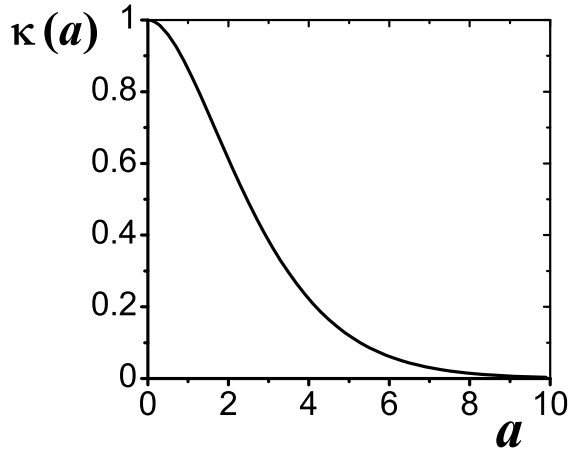,width=0.45\textwidth}
\caption{The suppression factor $\kappa(a)$ from
Eq.~(\ref{sigma1}) as a function of the parameter $a$.
\label{fig-kappa}}
\end{figure}

\vspace{0.2cm}
 As the first step,  we consider the suppression of
the $\sigma_{SB}$ constant. For this purpose the quasi-particle
approach of Ref.~\cite{GY} (see also recent papers \cite{rev} and
references therein) will be used. The system of interacting gluons
is treated as a gas of non-interacting quasiparticles with gluon
quantum numbers, but with mass $m(T)$ which depends on $T$. The
particle energy $\omega$ and momentum $k$ are assumed to be
connected as
%
%
$\omega = \left[k^{2}~+~m^{2}(T)\right]^{1/2}$~.
%
The
energy density and pressure take then the following form:
%
 \eq{
 \varepsilon(T)~&=~\frac{d}{2\pi^{2}}\int_{0}^{\infty}k^2 dk
~\frac{\omega}{\exp(\omega/T)~-~1}~
+~B^*(T)~\equiv~\varepsilon_0(T,\omega)~+~B^*(T)~,\label{epsilon}
\\
 p(T)~&=~\frac{d}{6\pi^{2}}~ \int_{0}^{\infty} k^2 dk
~\frac{k^{2}}{\omega}~\frac{1}{\exp(\omega/T)~-~1}
~-~B^*(T)~\equiv~p_0(T,\omega)~-~B^*(T)~,\label{pressure}
}
where the degeneracy factor $d=2(N_{c}^{2}-1)$ equals 16 for the
SU(3) gluodynamics. The temperature dependent function $B^*(T)$ in
Eq.~(\ref{epsilon}) was introduced for the first time
in Ref.~\cite{GY}. It results from the
thermodynamical relation,
\eq{\label{therm}
T\frac{dp}{dT}~-~p(T)~=~\varepsilon(T)~,
}
which leads to the equation for the function $B^*(T)$,
\eq{\label{BT}
 ~\frac{dB^*}{dT} ~=~ -~\frac{\Delta_0(T,\omega)}{m}~\frac{dm}{dT}~,
}
where $ \Delta_0\equiv \varepsilon_0~-~3p_0$, and $\varepsilon_0$,
$p_0$ defined by Eqs.~(\ref{epsilon},\ref{pressure}) are the ideal
gas expressions for massive bosons. If the function $m(T)$ is
known one can calculate $B^*(T)$ from Eq.~(\ref{BT}) up to an
arbitrary integration constant $B$.
%
%
%
%
%
%
The linear relation $m=aT$ with $a=\text{const}\ge 0$ used for all
$T\ge T_c$ guarantees the high temperature behavior of
$\varepsilon(T)$ and $p(T)$ in agreement with the MC lattice
results.
%
%
%
%
For $m=aT$, the function $B^*(T)$ derived from Eq.~(\ref{BT})
equals to
%
%
$B^*(T)=B-~\Delta_0(T,\omega)/4$~.
%
%
One obtains the energy  density (\ref{epsilon}) and the pressure
(\ref{pressure}),
%
\eq{\label{EOSaT}
 \varepsilon(T)~=~\sigma~T^4~+~B~,~~~~~~ p(T)~=~\frac{\sigma}{3}~T^4~-~B~,
}
where the modified SB constant $\sigma$ equals to:
 \eq{ \label{sigma1}
 \sigma~ = ~{3d\over 2\pi^{2}}
~\sum_{n=1}^{\infty}\left[ {a^{2}\over n^{2}}~K_2(na)~ +~{a^3\over
4n}~
 K_{1}(na)\right] ~\equiv~ \kappa(a)~\sigma_{SB}~.
}
The $K_{1}$ and $K_{2}$ in Eq.~(\ref{sigma1}) are the modified
Bessel functions. The constant $\sigma$ in Eq.~(\ref{EOSaT})
includes the suppression factor $\kappa(a)$ which is defined by
Eq.~(\ref{sigma1}) and presented in Fig.~\ref{fig-kappa}.
%
Therefore, an assumption of the linear $T$-dependent mass, $m=aT$,
leads to the EoS (\ref{EOSaT}) similar to the bag model EoS, but
with the suppressed SB constant (\ref{sigma1}). For $a\rightarrow
0$, $\kappa \rightarrow 1$ follows, and Eq.~(\ref{EOSaT})
coincides with Eq.~(\ref{bag}).  The modified SB constant $\sigma=
4.73<\sigma_{SB}$ allows to fit the high temperature behavior of
$\varepsilon(T)$ and $p(T)$. This requires $\kappa(a)\cong 0.90$
and $a\cong 0.84$.

\vspace{0.2cm}
At the second step, which is the main point of our model
construction, we include the linear in $T$ contribution to the QGP
pressure.
If the function $\varepsilon(T)$ is known, Eq.~(\ref{therm}) is a
$1^{\text{st}}$  order differential equation for the function
$p(T)$. The general solution of this equation includes an
arbitrary integration constant  which results in the linear in
temperature term in the function $p(T)$. This was discussed for
the first time
in Ref.~\cite{GM}. Thus, for $\varepsilon(T)$ in
the form of Eq.~(\ref{EOSaT}), the general solution of
Eq.~(\ref{therm}) for $p(T)$ can be written as follows,
%
 %
%
%
\eq{\label{epsilon1}
 \varepsilon(T)~=~\sigma~T^4~+~B~,~~~~~~
 p(T)~=~\frac{\sigma}{3}~T^4~-~B~-A~T~.
}
%
A sum of the first and second terms in the expression for $p(T)$
is a partial solution of the inhomogeneous differential equation
(\ref{therm}) with $\varepsilon(T)$ given by (\ref{epsilon1}),
whereas the last term in  $p(T)$  corresponds to a general
solution of the homogeneous equation $Tdp/dT - p=0$. Therefore,
the thermodynamical relation (\ref{therm}) between the pressure
and energy density  admits the linear in $T$ contribution to
$p(T)$, which is fully invisible in the $\varepsilon(T)$ function.
%
%
%
%

\begin{figure}[ht!]
\epsfig{file=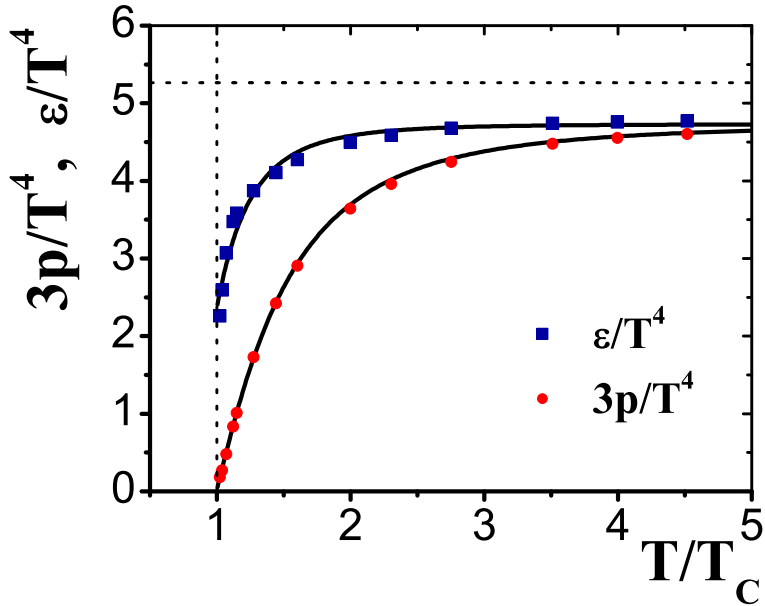,width=0.6\textwidth}
\caption{The MC lattice results and the dotted horizontal line are
the same as in Fig.~1. The solid lines correspond to the  $A$-BM
EoS (\ref{epsilon1}) with $\sigma=4.73$, $A=3.94~T_c^3$, and
$B=-~2.37~T_c^4$ for $\varepsilon/T^4$ (upper line) and $3p/T^4$
(lower line). \label{fig-AB} }
\end{figure}

Equation~(\ref{epsilon1}) defines our model suggestion for the QGP
EoS. For brevity we call it the $A$-bag model ($A$-BM). The
formula for $\varepsilon(T)$ looks formally the same as in
Eq.~(\ref{bag}). However, the pressure function $p(T)$ in the
$A$-BM (\ref{epsilon1}) contains one more parameter $A$ comparing
to the original BM EoS~(\ref{bag}).
The model parameters, $\sigma=4.73$
and $B=-~2.37~T_c^4$  are found from fitting the MC lattice
results \cite{lattice} for the energy density function
$\varepsilon(T)$.
The third $A$-BM parameter $A=3.94~T_c^3$ is fixed by fitting the
pressure function $p(T)$.
One finds a good description of the MC lattice results for
$\varepsilon(T)$ and $p(T)$ within the $A$-BM EoS (\ref{epsilon1})
for all~\footnote{To be precise, note that we consider the MC
lattice points with $T\ge 1.02~T_c$ to avoid the uncertainties of
$\varepsilon(T)$ at $T=T_c$~.  }~ $T>T_c$ as shown in
Fig.~\ref{fig-AB}. The parameter $\sigma$ in Eq.~(\ref{epsilon1})
regulates the high temperature behavior of $\varepsilon/T^4\cong
3p/T^4\cong \sigma$. As $A>0$, the linear in $T$ term gives a
negative contribution to $p(T)$ and guarantees both the correct
high temperature asymptotic behavior of $p(T)$ and its strong drop
at $T$ near $T_c$.
The bag parameter $B$ in Eq.~(\ref{epsilon1}) is found to be {\it
negative}, in contrast to the {\it positive} bag constant $B$ in
the standard BM EoS (\ref{bag}). Thus, according to the $A$-BM
(\ref{epsilon1}), $\varepsilon/T^4$ approaches its high
temperature limit $\sigma$ from below. This is in agreement with
the MC lattice results.
%

\vspace{0.2cm}
 An important characteristic of the EoS is the
so-called interaction measure, $(\varepsilon -3p)/T^4$, which
shows the deviation from the system of noninteracting massless
particles. For the $A$-BM EoS (\ref{epsilon1}) the interaction
measure reads,
 \eq{\label{delta}
\frac{\varepsilon ~-~3p}{T^4}~=~ \frac{3A}{T^3}~+~\frac{4B}{T^4}~.
}
%
The MC lattice results \cite{lattice} demonstrate a prominent
maximum of the function $(\varepsilon - 3p)/T^4$ at $T_{max}\cong
1.1~T_c$. The maximum of $(\varepsilon - 3p)/T^4$ is described in
the $A$-BM (\ref{epsilon1}). This happens due to different signs
of the A- and B-terms ($A>0$, $B<0$) in the r.h.s. of
Eq.~(\ref{delta}). Note that such a maximum is not reproduced by
the so-called fuzzy bag model \cite{pisarsky}. In that model,
there are $T^2$ contributions to both $p(T)$ and $\varepsilon(T)$,
\eq{\label{BC}
\varepsilon(T)~=~\sigma~T^4~-~C~T^2~+~B~,~~~~~~p(T)~=~
\frac{\sigma}{3}~T^4~-~C~T^2~-~B~,
}
and a comparison with the MC lattice results \cite{lattice} gives
$C>0$ and $B>0$. In that case, $(\varepsilon -3p)/T^4=
2C/T^2~+~4B/T^4$ corresponds to a monotonous decreasing function
of $T$ as both terms are positive. A comparison of the EoS
(\ref{epsilon1}) and (\ref{BC}) will be discussed in more details
in Ref.~\cite{BGM}. An extension of the $A$-BM to the SU$(N_c)$
gluodynamics with $N_c>3$ \cite{Nc}, to the quark degrees of
freedom and non-zero baryonic chemical potentials can be done
along the same scheme and will be considered elsewhere.


\vspace{0.2cm}
In summary, we have suggested a new EoS for the deconfined matter
-- the $A$-BM (\ref{epsilon1}). It satisfies all qualitative
features of the MC lattice results at $T>T_c$ and gives a good
quantitative description of the lattice results \cite{lattice} for
the SU(3) gluodynamics, see Fig.~\ref{fig-AB}. The expression for
$\varepsilon(T)$ in the $A$-BM (\ref{epsilon1}) looks similar to
that in the BM (\ref{bag}). However, the pressure function $p(T)$
in the $A$-BM (\ref{epsilon1}) contains a new linear in $T$
negative term which does not contribute to $\varepsilon(T)$. The
presence of this negative pressure term leads to a principal
difference between the bag term $B$ in the BM and that in the
$A$-BM. The bag parameter in the $A$-BM (\ref{epsilon1}) is found
to be {\it negative}, in contrast to the {\it positive} bag
constant $B$ in the BM EoS (\ref{bag}). The $A$-BM
(\ref{epsilon1}) gives a simple analytical parametrization of the
QGP EoS. This opens  new possibilities for its applications in the
hydrodynamic description of the QGP.

\vspace{0.2cm} {\bf Acknowledgments.} We would like to thank
A.I.~Bugrij, M.~Ga\'zdzicki, W.~Greiner, V.P.~Gusynin,
L.L.~Jenkovszky, and O.~Linnyk for fruitful discussions and
O.~Kaczmarek for providing us with the lattice results.  V.V.
Begun thanks the Alexander von Humboldt Foundation for the
support. This work was in part supported by the Program of
Fundamental Research of the Department of Physics and Astronomy of
NAS, Ukraine.



%

\end{document}